\documentclass[11pt]{revtex4}
\usepackage{amssymb}
\usepackage{latexsym}
\usepackage{epsfig}                                                 
\begin{document}
\title{Dilaton black holes coupled to nonlinear electrodynamic field}
\author{A. Sheykhi$^{1,2}$\footnote{asheykhi@shirazu.ac.ir},  S. Hajkhalili $^{1}$}
\address{$^1$ Physics Department and Biruni Observatory, College of
Sciences, Shiraz University, Shiraz 71454, Iran\\
$^2$ Research Institute for Astronomy and Astrophysics of Maragha
(RIAAM), P.O. Box 55134-441, Maragha, Iran}
\begin{abstract}
The theory of nonlinear electrodynamics has got a lot of
attentions in recent years. It was shown that Born-Infeld
nonlinear electrodynamics is not the only modification of the
linear Maxwell's field which keeps the electric field of a charged
point particle finite at the origin, and other type of nonlinear
Lagrangian such as exponential and logarithmic nonlinear
electrodynamics can play the same role. In this paper, we
generalize the study on the exponential nonlinear electrodynamics
by adding a scalar dilaton field to the action. By suitably
choosing the coupling of the matter field to the dilaton field, we
vary the action and obtain the corresponding field equations.
Then, by making a proper ansatz, we construct a new class of
charged dilaton black hole solutions coupled to the exponential
nonlinear electrodynamics field in the presence of two
Liouville-type potentials for the dilaton field. Due to the
presence of the dilaton field, the asymptotic behavior of these
solutions are neither flat nor (A)dS. In the limiting case where
the nonlinear parameter $\beta^2$ goes to infinity, our solution
reduces to the Einstein-Maxwell dilaton black holes. We obtain the
mass, temperature, entropy and electric potential of these
solutions. We also study the behaviour of the electric field as
well as the electric potential of these black holes near the
origin. We find that the electric field has a finite value
\textit{near} the origin, which is the same as the electric field
of Born-Infeld nonlinear electrodynamics, but it can diverge
exactly at $r=0$ depending on the model parameters.  We also
investigate the effects of the dilaton field on the behaviour of
the electric field and electric potential. Finally, we check the
validity of the first law of black hole thermodynamics on the
black hole horizon.

\end{abstract}

 \maketitle

\section{Introduction}
One of the challenges of the classical Maxwell theory is that it
leads to an infinite electric field at $r=0$, and hence an
infinite self energy for a charged point particle located at
$r=0$. Although, applying quantum electrodynamics one can remove
the divergences in the theory by using the renormalization
procedure, the problem is still remains in the classical
electrodynamics. The first attempt to resolve the divergence
problem appearing in the Maxwell theory was made by Born and
Infeld in $1934$ \cite{BI}. The Lagrangian of the Born-Infeld (BI)
nonlinear gauge field can be written \cite{BI}
\begin{equation}\label{LBI}
L_{\rm BI}=4\beta^{2}
\left(1-\sqrt{1+\frac{F^2}{2\beta^2}}\right),
\end{equation}
where $\beta $ is called the nonlinear parameter with dimension of
mass, $F^2=F_{\mu \nu }F^{\mu \nu }$, where $F_{\mu \nu }$ is the
electromagnetic field tensor. It has been shown that charged black
hole solutions in BI theory are less singular in comparison with
the Reissner-Nordstr\"{o}m solution. Indeed, in BI theory there is
no divergence term $q^{2}/r^{2}$ in the metric near the
singularity while the Schwarzschild-type term $m/r$ do
exit\cite{Car,Banad}. In recent years, BI nonlinear
electrodynamics has got a new impetus, since it naturally arises
in the low energy limit of the open string theory \cite{Frad,Cal}.
In addition, most of physical systems in the nature, including the
field equations of the gravitational systems, are intrinsically
nonlinear and hence including a nonlinear electrodynamics in the
action of the gravitational systems is well motivated. Black
object solutions coupled to a nonlinear BI gauge field has been
studied and their thermodynamics have been investigated in ample
details \cite{Fern,Tamaki,Dey,Cai1,MHD}.

In recent years, other types of nonlinear electrodynamics in the
context of gravitational field have been introduced, which can
also remove the divergence of the electric field at $r = 0$,
similar to BI nonlinear electrodynamics. Two well-known nonlinear
Lagrangian for electrodynamics are logarithmic and exponential
Lagrangian. Logarithmic nonlinear (LN) $U(1)$ gauge theory was
proposed by Soleng \cite{Soleng},
\begin{equation}\label{LLN}
L_{\rm LN}=-4\beta^{2} \ln\left(1+\frac{F^2}{4\beta^2}\right),
\end{equation}
while the Lagrangian of the exponential nonlinear (EN)
electrodynamics was suggested by Hendi as \cite{HendiJHEP},
\begin{equation}\label{LEN}
L_{\rm EN}=4\beta^{2}
\left[\exp\left(-\frac{F^2}{4\beta^2}\right)-1\right].
\end{equation}
It is worth mentioning that logarithmic form of the electrodynamic
Lagrangian, like BI electrodynamics, removes divergences in the
electric field, while the exponential form of nonlinear
electromagnetic field does not cancel the divergency of the
electric field at $r = 0$, however, its singularity is much weaker
than Einstein-Maxwell theory. Although these two type of nonlinear
electrodynamics have no direct relation to superstring theory,
they serve as a toy model illustrating that certain nonlinear
field theories can produce particle-like solutions which can
realize the limiting curvature hypothesis also for gauge fields
\cite{Soleng}. Besides, the expansions of these Lagrangians, for
large value of nonlinear parameter $\beta^2$, leads to Maxwell
linear Lagrangian, exactly like the BI case,
\begin{equation}\label{L}
L_{\rm BI}=L_{\rm LN}=L_{\rm EN}=-F^2+\frac{F^4}{8
\beta^2}+O\left(\frac{1}{\beta^4}\right).
\end{equation}
Clearly, for ${\beta^2\rightarrow \infty}$, we arrive at $L_{\rm
BI}=L_{\rm LN}=L_{\rm EN}=-F^2$. In addition, from the AdS/CFT
correspondence viewpoint in hydrodynamic models, it has been shown
that, unlike gravitational correction, higher-derivative terms for
Abelian fields in the form of nonlinear electrodynamics do not
affect the ratio of shear viscosity over entropy density
\cite{Caishear}. Furthermore, in applications of the AdS/CFT
correspondence to superconductivity, nonlinear electrodynamic
theories make crucial effects on the condensation  as well as the
critical temperature of the superconductor and its energy gap
\cite{Jing,Greg}.

For all mentioned above, further studies on EN electrodynamics are
well motivated. It is also interesting to extend the study to the
dilaton gravity. Of particular interest is to investigate the
effects of the dilaton field on the physical properties of the
solutions. The appearance of the dilaton field changes the
asymptotic behavior of the solutions to be neither asymptotically
flat nor (A)dS. There are at least two motivations for
investigating non-asymptotically flat, non-asymptotically (A)dS
black hole spacetimes. First, these kind of solutions might lead
to possible extensions of AdS/CFT correspondence. Indeed, it has
been speculated that the linear dilaton spacetimes, which arise as
near-horizon limits of dilatonic black holes, might exhibit
holography \cite{Ahar}. Second, such kind of solutions may be used
to extend the range of validity of methods and tools originally
developed for, and tested in the case of, asymptotically flat or
asymptotically (A)dS black holes. The BI action including a
dilaton and an axion field, appears in the coupling of an open
superstring and an Abelian gauge field theory \cite{Frad}. This
action, describing a Born-Infeld-dilaton-axion system coupled to
Einstein gravity, can be considered as a nonlinear extension in
the Abelian field of Einstein-Maxwell-dilaton-axion gravity.
Although one can consistently truncate such models, the presence
of the dilaton field cannot be ignored if one considers coupling
of the gravity to other gauge fields, and therefore one remains
with Einstein-BI gravity in the presence of a dilaton field.
Physical properties, thermodynamics and thermal stability of the
black object solutions in the context of BI dilaton theory have
been investigated \cite{Tam1,Tam2,YI,yaz,
Clement,yazad,SRM,Shey1,Shey2,DHSR}.

In  this paper, we turn the investigation on EN electrodynamics by
including a dilaton field in the action. When the dilaton field is
coupled to the EN gauge field, it has profound consequences for
the black hole solutions. Due to the presence of the dilaton
field, the asymptotic behaviour of these solutions are neither
flat nor (A)dS. We shall investigate the effects of the
nonlinearity as well as the dilaton field on the properties of the
solutions. We also compute the conserved and thermodynamic
quantities of these solutions and check the validity of the first
law of thermodynamics on the black hole horizon.

The organization of this paper is as follows. In Sec. \ref{Field},
we introduce the Lagrangian of EN electrodynamics coupled to the
dilaton field in Einstein gravity, and obtain the corresponding
field equations by varying the action. Then, we construct a new
class of black hole solutions in the presence of two Liouville
type potentials for the dilaton field and general dilaton coupling
constant. In Sec. \ref{Pro}, we investigate the physical
properties of the obtained solutions. In Sec. \ref{Therm}, we
obtain the conserved and thermodynamic quantities of the spacetime
and show that these quantities satisfy the first law of
thermodynamics. We finish our paper with some concluding remarks.

\section{Basic Equations and Solutions}\label{Field}
Our starting point is the  following action in which gravity is
coupled to dilaton and nonlinear electrodynamic fields
\begin{equation}\label{Act}
S=\frac{1}{16\pi}\int{d^{4}x\sqrt{-g}\left(\mathcal{R}\text{ }-2
g^{\mu\nu}
\partial_{\mu} \Phi \partial_{\nu}\Phi -V(\Phi
)+L(F,\Phi)\right)},
\end{equation}
where $\mathcal{R}$ and  $\Phi $ are, respectively, the Ricci
scalar curvature and the dilaton filed, and $V(\Phi )$ is the
potential for $\Phi $. We choose the Lagrangian of the EN
electrodynamics coupled to the dilaton field (ENd), $L(F,\Phi)$,
as
\begin{equation}\label{LFPhi}
L(F,\Phi)=4\beta^{2} e^{2\alpha \Phi}\left[\exp\left(-\frac{e^{-
4\alpha \Phi}F^2}{4\beta^{2}}\right)-1\right],
\end{equation}
where $\alpha$ is a constant determining the strength of coupling
of the scalar and electromagnetic fields. In order to motivate
such a choice for the Lagrangian of ENd field, let us invoke the
BI-dilaton (BId) Lagrangian which is written as \cite{Tam1}
\begin{equation}\label{LFPhiB}
L_{\rm BId}(F,\Phi)=4\beta^{2} e^{2\alpha \Phi}\left(
1-\sqrt{1+\frac{e^{- 4\alpha \Phi}F^2}{2\beta^{2}}}\right).
\end{equation}
Note that Lagrangian (\ref{LFPhiB}) originates from the open
string version of the BI action coupled to a dilaton field and
only valid for the pure electric case \cite{Tam1}. Clearly, this
version of the BId action does not enjoy electric-magnetic duality
\cite{Clement}. This form for the BId term have been previously
investigated by a number of authors \cite{Tam1,Tam2,YI,yaz,
Clement,yazad,SRM,Shey1,Shey2,DHSR}. Expanding the BId Lagrangian
(\ref{LFPhiB}) for large $\beta$, leads to
\begin{equation}\label{ExB}
L_{\rm BId}(F,\Phi)= -e^{-2\alpha \Phi}F^2+\frac{ e^{-6\alpha
\Phi}F^4}{8\beta^{2}}-\frac{ e^{-10\alpha
\Phi}F^6}{32\beta^{4}}+O\left(\frac{1}{\beta^6}\right),
\end{equation}
while the expansion of (\ref{LFPhi}) for large $\beta$, yields
\begin{equation}\label{ExE}
L(F,\Phi)=- e^{-2\alpha \Phi}F^2+\frac{ e^{-6\alpha
\Phi}F^4}{8\beta^{2}}-\frac{ e^{-10\alpha
\Phi}F^6}{96\beta^{4}}+O\left(\frac{1}{\beta^6}\right).
\end{equation}
As one can see, the BId Lagrangian and the ENd Lagrangian given in
(\ref{LFPhi}) have similar expansions and so it is interesting to
replace (\ref{LFPhiB}) with (\ref{LFPhi}) in the action, and
investigate the effects of this kind of nonlinear electrodynamics
coupled to the dilaton field on the behavior of the solutions. In
the absence of the dilaton field ($\alpha=0$), $L(F,\Phi)$ reduces
to (\ref{LEN}) as expected. On the other hand, in the limiting
case $\beta^2 \rightarrow \infty $, both $L(F,\Phi)$ and $L_{\rm
BId}(F,\Phi)$ recovers the standard linear Maxwell Lagrangian
coupled to the dilaton field \cite{CHM}
\begin{equation}
L(F,\Phi)=L_{\rm BId}(F,\Phi)=-e^{-2\alpha \Phi }F^2.
\end{equation}
It is convenient to set
\begin{equation}
L(F,\Phi)=4\beta^2 e^{2\alpha \Phi}{\mathcal{L}}(Y),
\end{equation}
where we define
\begin{eqnarray}
{\mathcal{L}}(Y) &=&\exp(-Y)-1,\label{LY}\\
Y&=& \frac{e^{-4\alpha \Phi}F^2}{4\beta^2}.\label{Y}
\end{eqnarray}
In order to derive the field equations, we vary action (\ref{Act})
with respect to the gravitational field $g_{\mu \nu }$, the
dilaton field $\Phi $ and the electromagnetic field $A_{\mu }$. We
find
\begin{eqnarray}\label{FE1}
{\cal R}_{\mu\nu}&=& 2 \partial _{\mu }\Phi
\partial _{\nu }\Phi +\frac{1}{2}g_{\mu \nu }V(\Phi)-
2e^{-2\alpha \Phi}\partial_{Y}{{\cal L}}(Y) F_{\mu\eta}
F_{\nu}^{\text{ }\eta } \nonumber \\
&&+2\beta^2 e^{2\alpha \Phi} \left[2Y\partial_{Y}{{\cal
L}}(Y)-{{\cal L}}(Y)\right]g_{\mu\nu},
\end{eqnarray}
\begin{equation}\label{FE2}
\nabla ^{2}\Phi =\frac{1}{4}\frac{\partial V}{\partial \Phi}+
2\alpha \beta^2 e^{2\alpha \Phi }\left[2{ Y}\partial_{Y}{{\cal
L}}(Y)-{\cal L}(Y)\right],
\end{equation}
\begin{equation}\label{FE3}
\nabla _{\mu }\left(e^{-2\alpha \Phi}
\partial_{Y}{{\cal L}}(Y) F^{\mu\nu}\right)=0.
\end{equation}
In case of linear electrodynamics we have ${\cal L}(Y)=-Y$, and
the system of equations (\ref{FE1})-(\ref{FE3}) reduce to the
well-known equations of Einstein-Maxwell-dilaton (EMd) gravity
\cite{CHM,Cai,Clem,Shey3}.

In the present work, we search for the static and spherically
symmetric black hole solutions of the above field equations. We
assume the spacetime metric has the following form
\begin{equation}\label{metric}
ds^2=-f(r)dt^2 +{dr^2\over f(r)}+
r^2R^2(r)\left(d\theta^2+\sin^2\theta d\phi^2\right) ,
\end{equation}
where $f(r)$ and $R(r)$ are functions of $r$ which should be
determined. The additional function $R(r)$ is introduced in the
metric due to the presence of the new degree of freedom associated
with the dilation field in the equations of motion. First of all,
we integrate the electromagnetic field equation (\ref{FE3}). For
this purpose we assume all the components of $F_{\mu\nu}$ are zero
except $ F_{tr}$:
\begin{equation}\label{FtrE}
F_{tr}=\frac{q e^{2\alpha \Phi}}{r^2 R^2(r)}
\exp\left[-\frac{1}{2}L_W\left(\frac{q^2}{\beta^2 r^4
R^4(r)}\right)\right],
\end{equation}
where $q$ is an integration constant which is related to the
electric charge of the black hole, and $L_W(x)={LambertW(x)}$ is
the Lambert function which satisfies \cite{Lambert}
\begin{equation}
L_W(x)e^{L_W(x)}=x,
\end{equation}
and has the following series expansion
\begin{equation}\label{LW}
L_W(x)=x-x^2+\frac{3}{2}x^3-\frac{8}{3}x^4+....
\end{equation}
Clearly, series (\ref{LW}) converges for $|x| <1$.  Let us note
that the electric field in case of BId black holes is given by
\cite{Shey2}
\begin{equation}\label{FtrB}
F^{\rm BId}_{tr}=\frac{q e^{2\alpha \Phi}}{r^2
R^2(r)}\left[{1+\frac{q^{2}}{\beta^2 r^4 R^4(r)}}\right]^{-1/2}.
\end{equation}
In the absence of the dilaton field where $\alpha=0$ and $R(r)=1$,
Eq. (\ref{FtrE}) reduces to
\begin{equation}\label{FtrE2}
F_{tr}=\frac{q }{r^2}
\exp\left[-\frac{1}{2}L_W\left(\frac{q^2}{\beta^2 r^4
}\right)\right],
\end{equation}
while in the limiting case where $\beta \rightarrow \infty $, both
$Ftr$ and $F^{\rm BId}_{tr}$ reduces to the electric field of EMd
black holes \cite{Shey3}
\begin{equation}
F_{tr}=\frac{q e^{2\alpha \Phi}}{r^2 R^2(r)}.  \label{FtrM}
\end{equation}
The expansion of (\ref{FtrE}) for large $\beta$ is given by
\begin{equation}\label{FtrExp}
Ftr=\frac{q e^{2\alpha \Phi}}{r^2 R^2(r)}-\frac{1}{2
\beta^2}\frac{q^3 e^{2\alpha \Phi}}{r^6
R^6(r)}+O\left(\frac{1}{\beta^{4}}\right).
\end{equation}
In order to construct exact analytical solutions of the field
equation (\ref{FE1}) and (\ref{FE2}) for an arbitrary dilaton
coupling constant $\alpha$, we assume the dilaton potential
contain two Liouville terms,
\begin{equation}\label{v2}
V(\Phi) = 2\Lambda_{0} e^{2\zeta_{0}\Phi} +2 \Lambda e^{2\zeta
\Phi},
\end{equation}
where $\Lambda_{0}$,  $\Lambda$, $ \zeta_{0}$ and $ \zeta$ are
constants. This kind of potential was previously investigated in
the context of EMd gravity \cite{CHM,Shey3} as well as BId black
holes \cite{Shey1,Shey2}. The system of equations (\ref{FE1}) and
(\ref{FE2}) contain three unknown functions $f(r)$, $R(r)$ and
$\Phi (r)$. In order to solve them, we make the ansatz
\begin{equation}
R(r)=e^{\alpha \Phi}.\label{Rphi}
\end{equation}
This ansatz was first introduced in \cite{DF} for the purpose of
finding black string solutions of EMd gravity, and latter was
applied for constructing black hole solutions of nonlinear BId
theory \cite{Shey2}. It is important to note that in the absence
of the dilaton field ($\alpha=0$), we have $R(r)=1$, as one
expected (see Eq. \ref{metric}). Substituting (\ref{Rphi}), the
electromagnetic field (\ref{FtrE}) and the metric (\ref{metric})
into the field equations (\ref{FE1}) and (\ref{FE2}), one can
obtain the following solutions
\begin{eqnarray}
f(r)&=&-\frac{ {\alpha}^{2}+1}{ {\alpha}^{2}-1
}{b}^{-\gamma}{r}^{\gamma}-\frac{m}{r^{1-\gamma}}+\frac{(\Lambda+2\beta^2)
\left( {\alpha}^{2}+1 \right) ^{2}{b}^{\gamma}}{\alpha^{2}-3
}r^{2-\gamma}\nonumber \\
&&-\frac{2\beta q} {r^{1-\gamma}} \left( \alpha ^{2}+1\right)
b^{\gamma }\int r
^{-\gamma}\left(\sqrt{L_{W}(\eta)}-\frac{1}{\sqrt{L_{W}(\eta)}}\right){dr},
\label{f1}
\end{eqnarray}
\begin{equation}\label{phi}
\Phi (r)=\frac{\alpha }{\alpha ^{2}+1}\ln
\left(\frac{b}{r}\right),
\end{equation}
where $b$ is an arbitrary constant, $\gamma =2\alpha
^{2}/(1+\alpha ^{2})$, and
\begin{equation}
\eta\equiv \frac{q^{2}r^{2\gamma-4}}{\beta ^{2}b^{2\gamma }}.
\label{eta}
\end{equation}
In the above expression, $m$ appears as an integration constant
and is related to the Arnowitt-Deser-Misner (ADM) mass of the
black hole. The obtained solutions fully satisfy the system of
equations (\ref{FE1}) and (\ref{FE2}) provided we take
\begin{eqnarray}
\zeta_{0} =\frac{1}{\alpha},  \ \ \ \zeta=\alpha, \ \ \
\Lambda_{0}= \frac{ b^{-2}\alpha^2 }{\alpha^2-1}.
\end{eqnarray}
Notice that $\Lambda$ remains as a free parameter which plays the
role of the cosmological constant. One can redefine it as
$\Lambda=-3/l^2$, where $l$ is a constant with dimension of
length.  The integral of Eq. (\ref{f1}) can be performed using the
Mathematica software. The resulting solution can be written
\begin{eqnarray}\label{f2}
f(r)&=&-\frac{ {\alpha}^{2}+1}{ {\alpha}^{2}-1
}{b}^{-\gamma}{r}^{\gamma}-\frac{m}{r^{1-\gamma}}+\frac{(\Lambda+2\beta^2)
\left( {\alpha}^{2}+1 \right) ^{2}{b}^{\gamma}}{\alpha^{2}-3
}r^{2-\gamma}+\frac{\beta q (\alpha^2+1)^4}{2
(\alpha^2-1)^2}\left(\frac{\beta^2
b^{2\gamma}}{q^2}\right)^{\frac{1-\gamma}{2\gamma-4}}r^{\gamma-1}
\nonumber
\\ &&\times
\left(\frac{1-\alpha^2}{4}\right)^{\frac{2\gamma-3}{2\gamma-4}}
\Bigg{\{} -4(\gamma-2)^2\left[\Gamma
\left(\frac{\alpha^2+5}{4},\frac{1-\alpha^2}{4}L_{W}(\eta)\right)-\Gamma\left({\frac{\alpha^2+5}{4}}\right)\right]\nonumber \\
&&+(\gamma-1)^2\left[\Gamma
\left(\frac{\alpha^2-3}{4},\frac{1-\alpha^2}{4}L_{W}(\eta)\right)
-\Gamma\left({\frac{\alpha^2-3}{4}}\right)\right]\Bigg{\}}.
\end{eqnarray}
where $\Gamma(a,z)$ and $\Gamma(a)$ are Gamma functions and they
are related to each other as,
\begin{eqnarray}\label{Gamma}
\Gamma(a,z)=\Gamma(a)-\frac{z^a }{a}\digamma\left(a,1+a,-z\right).
\end{eqnarray}
where $ \digamma(a,b,z)$ is hypergeometric function
\cite{Lambert}. Using (\ref{Gamma}), solution (\ref{f2}) can also
be expressed in terms of hypergeometric function,
\begin{eqnarray}\label{fhyper}
f(r)&=&-\frac{ {\alpha}^{2}+1}{ {\alpha}^{2}-1
}{b}^{-\gamma}{r}^{\gamma}-\frac{m}{r^{1-\gamma}}+\frac{(\Lambda+2\beta^2)
\left( {\alpha}^{2}+1 \right) ^{2}{b}^{\gamma}}{\alpha^{2}-3
}r^{2-\gamma}+2 \beta q{(\alpha^2+1)^2}
r^{\gamma-1}{(\frac{\beta^2
b^{2\gamma}}{q^2})}^{\frac{1-\gamma}{2\gamma-4}}{L_W^{\frac{3-2\gamma}{2\gamma-4}}(\eta)}
\nonumber
\\ &&\times\Bigg{\{}\frac{L_W^2(\eta)}{\alpha^2+5}\digamma\left(\frac{\alpha^2+5}{4},\frac{\alpha^2+9}{4},\frac{\alpha^2-1}{4}L_W(\eta)\right)
-
\frac{1}{\alpha^2-3}\digamma\left(\frac{\alpha^2-3}{4},\frac{\alpha^2+1}{4},\frac{\alpha^2-1}{4}L_W(\eta)\right)
\Bigg{\}}. \nonumber\\
\end{eqnarray}
In the absence of the dilaton field ($\alpha=0=\gamma$), our
solution reduces to the one obtained in \cite{Hendi2} for
asymptotic Reissner-Nordstr\"{o}m (RN) black hole coupled to EN
electrodynamics in AdS spaces. Using the fact that $L_W(x)$ has a
convergent series expansion for $|x| <1$ as given in (\ref{LW}),
we can expand (\ref{f2}) for large $\beta$. We find
\begin{eqnarray}\label{f3}
f(r)&=&-\frac{\alpha^2+1}{\alpha^2-1}b^{-\gamma}
r^\gamma-\frac{m}{r^{1-\gamma}}+\frac{\Lambda(\alpha^2+1)^2}{\alpha^2-3}
b^\gamma r^{2-\gamma}\nonumber\\
&&+\frac{({\alpha^2+1})b^{-\gamma}q^2}{ r^{2-\gamma}}-\frac{q^4
(\alpha^2+1)^2b^{-3\gamma}}{4\beta^2(\alpha^2+5)r^{6-3\gamma}}+O\left(\frac{1}{\beta^4}\right).
\end{eqnarray}
This is exactly the result obtained in \cite{Shey2} for BId black
holes in the limit of large $\beta$. This is an expected result,
since as we discussed already, in the limit of large $\beta$, the
Lagrangian of BId and Lagrangian of ENd electrodynamics have the
same expansion and thus the resulting solutions have the same
behavior too. In the absence of the dilaton field ($\alpha
=0=\gamma$), solution (\ref{f3}) reduces to
\begin{eqnarray}\label{f4}
f(r) &=&1-\frac{m}{r}-\frac{\Lambda}{3} r^2+\frac{q^{2}}{r^{2}}-
\frac{1}{20\beta^2}\frac{q^{4}}{r^{6}}+O\left(\frac{1}{\beta^4}\right),
\end{eqnarray}
which has the form of static spherically symmetric RN black hole
in AdS spacetime in the limit $\beta\rightarrow \infty$. The last
term in the right hand side of (\ref{f4}) is the leading nonlinear
correction term to the RN-AdS black hole in the large $\beta$
limit.
\section{Physical properties of solutions}\label{Pro}
\begin{figure}[tbp]
\epsfxsize=7cm \centerline{\epsffile{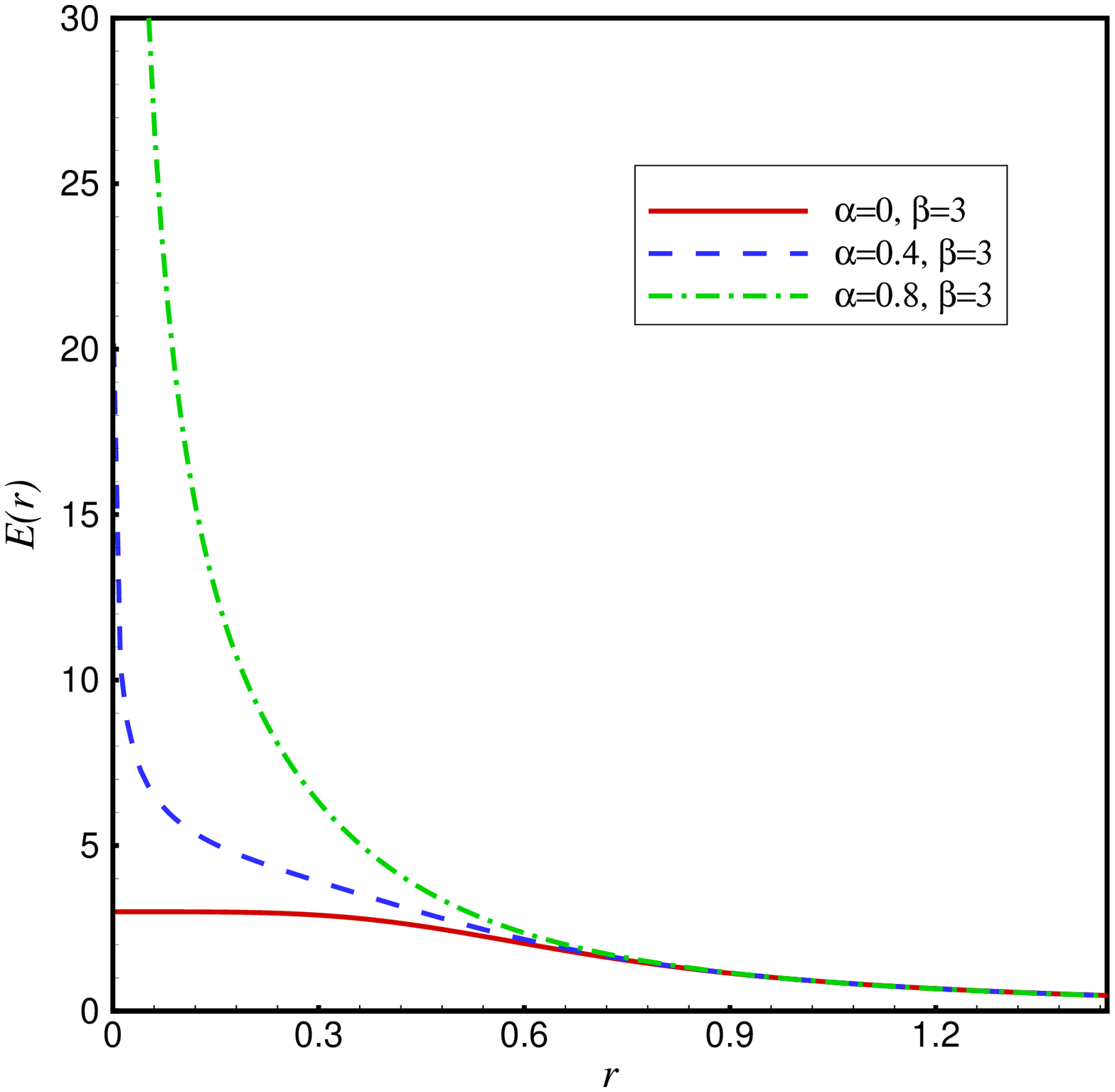}} \caption{The
behavior of the electric field $E(r)$ of BId black holes versus
$r$ for $b=1$ and $q=1$.} \label{figure1}
\end{figure}

\begin{figure}[tbp]
\epsfxsize=7cm \centerline{\epsffile{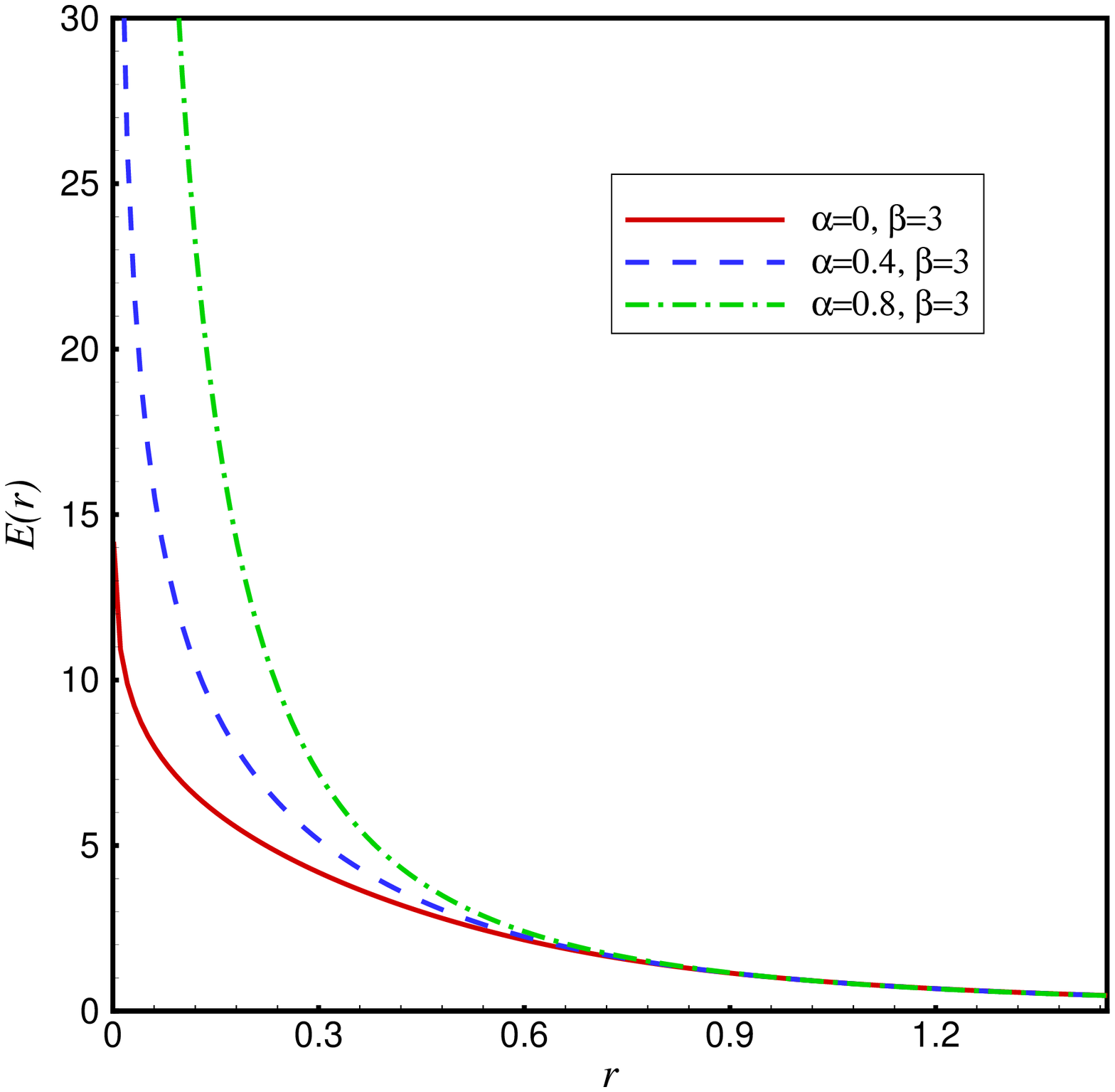}} \caption{The
behavior of the electric field $E(r)$ of ENd black holes versus
$r$ for $b=1$ and $q=1$.} \label{figure2}
\end{figure}

\begin{figure}[tbp]
\epsfxsize=7cm \centerline{\epsffile{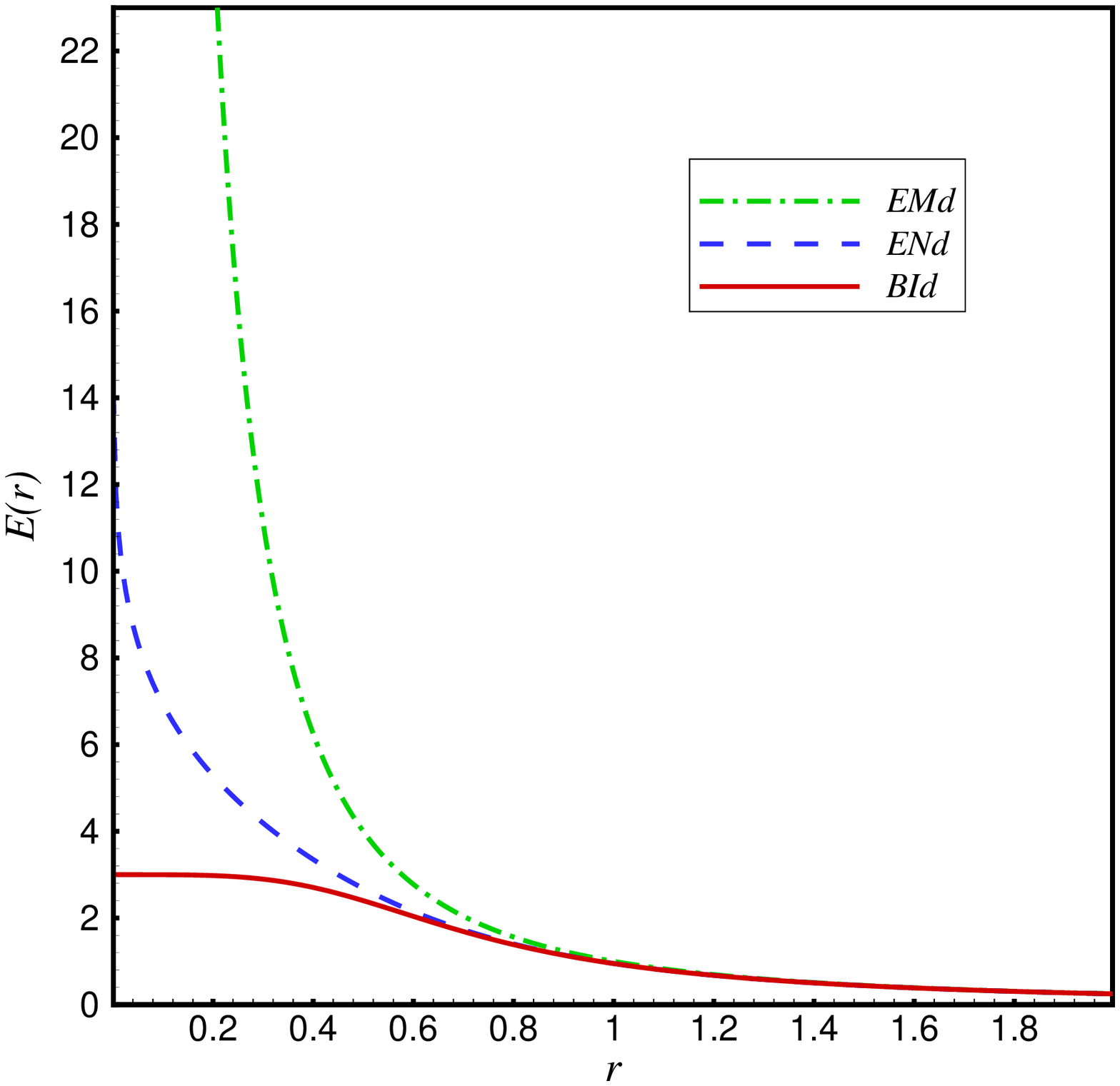}} \caption{The
behavior of the electric field $E(r)$ versus $r$ for $b=1$,
$\beta=3$, $\protect\alpha=0$ and $q=1$.} \label{figure3}
\end{figure}

\begin{figure}[tbp]
\epsfxsize=7cm \centerline{\epsffile{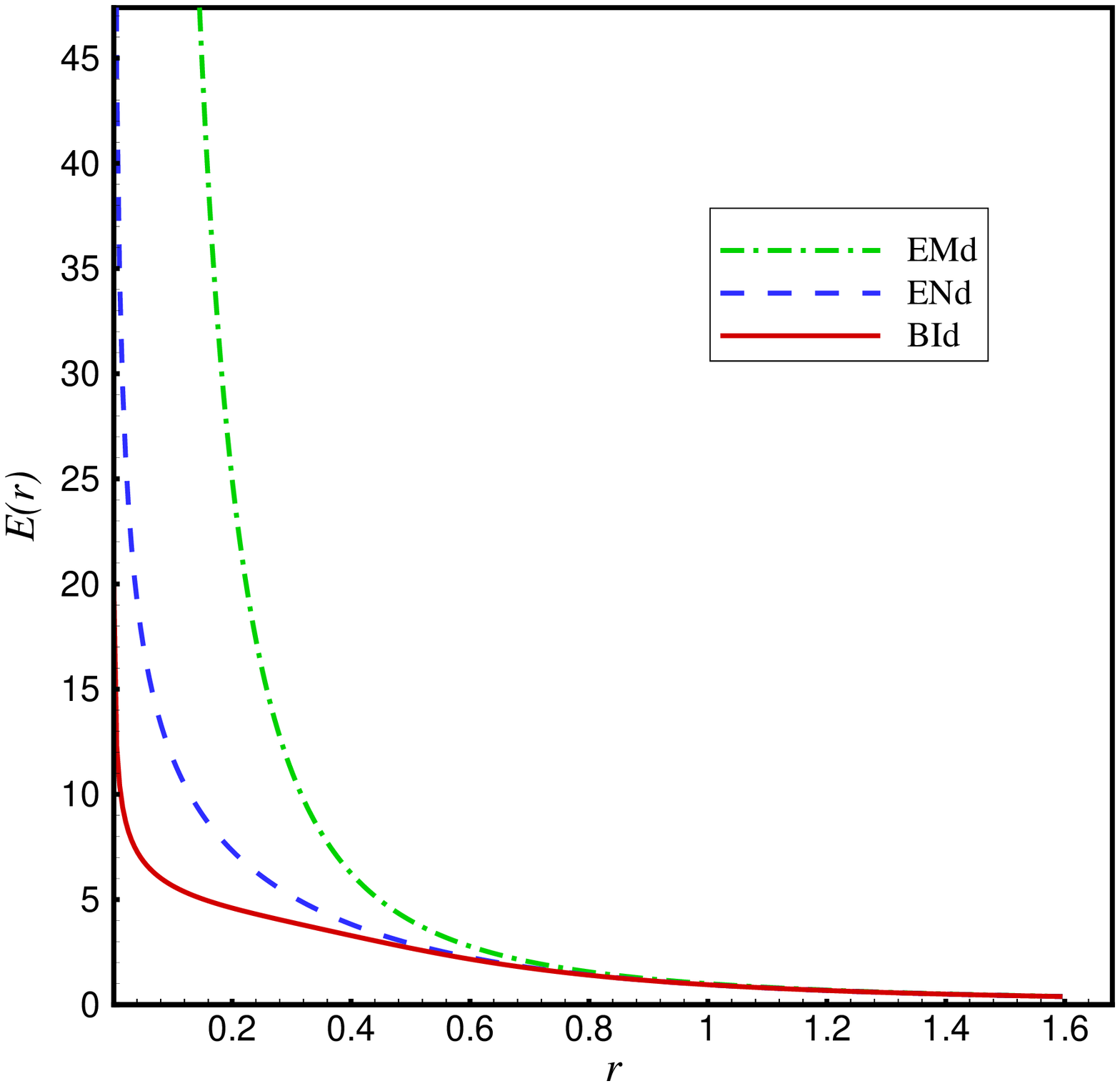}} \caption{The
behavior of the electric field $E(r)$ versus $r$ for $b=1$,
$\beta=3$, $\protect\alpha=0.4$ and $q=1$.} \label{figure4}
\end{figure}

\begin{figure}[tbp]
\epsfxsize=7cm \centerline{\epsffile{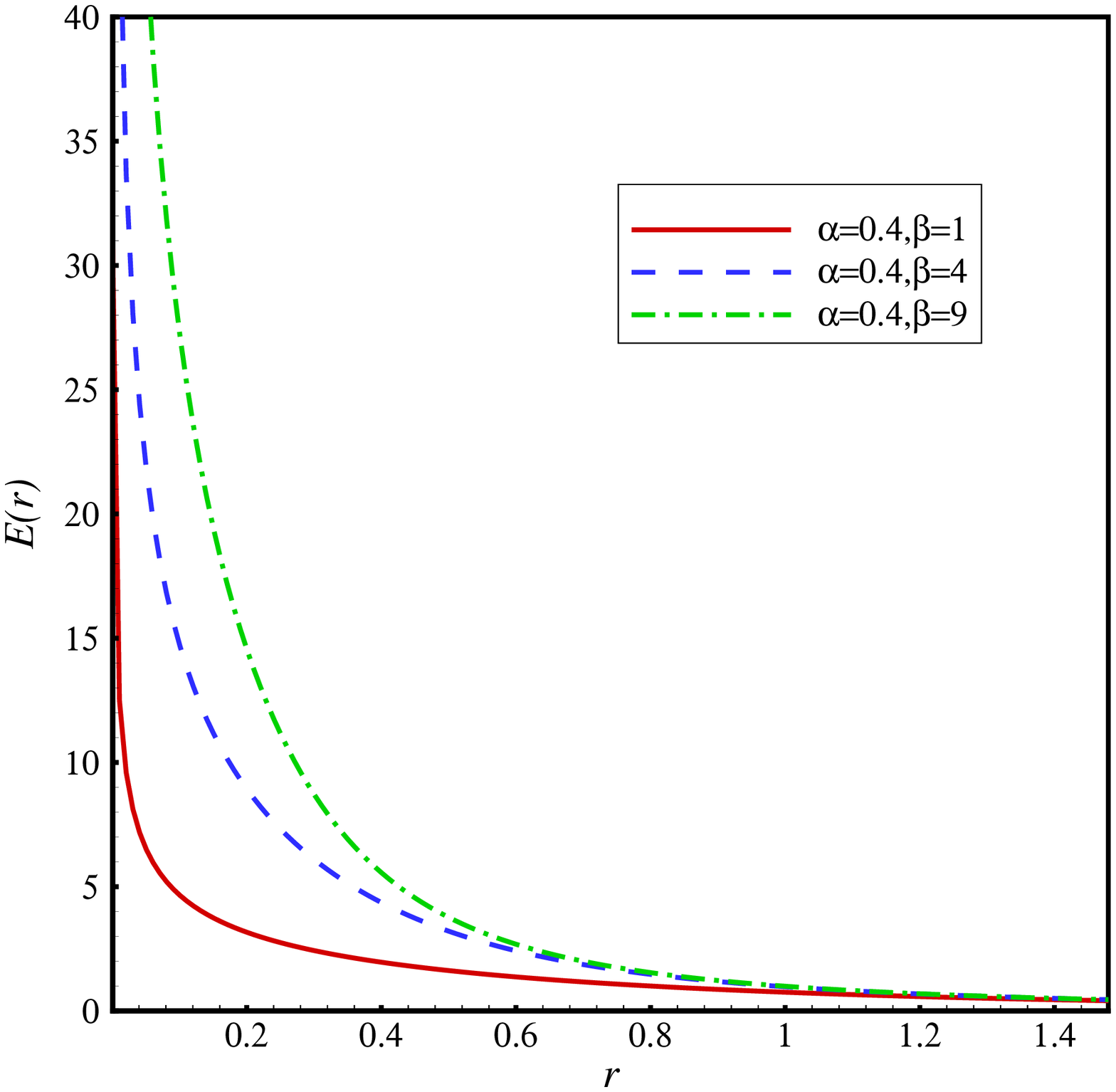}} \caption{The
behavior of the electric field $E(r)$ of ENd black holes versus
$r$ for $b=1$ and $q=1$.} \label{figure5}
\end{figure}

In this section we would like to investigate the physical
properties of the solutions. For this purpose, we first study the
behavior of the electric field of the obtained solution. Combining
Eqs. (\ref{Rphi}) and (\ref{phi}) with (\ref{FtrE}), we find
\begin{equation}\label{FtrE2}
F_{tr}=\frac{q}{r^2}
\exp\left[-\frac{1}{2}L_W\left(\frac{q^{2}r^{2\gamma-4}}{\beta
^{2}b^{2\gamma }}\right)\right].
\end{equation}
Expanding for large $\beta$, we arrive at
\begin{equation}\label{FtrE2}
F_{tr}=\frac{q}{r^2} -\frac{q^3 b^{-2\gamma }}{2 \beta^2
r^{6-2\gamma}}+O\left(\frac{1}{\beta^4}\right).
\end{equation}
We have plotted the behavior of the electric field versus $r$ in
Figs. \ref{figure1}-\ref{figure5}. From these figures we see that
in all cases the electric field goes to zero for large $r$
independent of the value of the other parameters. Figure 1 shows
that for BId black holes, and in the absence of the dilaton field
($\alpha=0$), the electric field has a finite value at $r=0$,
while as soon as the dilaton field is taken into account
($\alpha>0$), the electric field diverges as $r\rightarrow 0$. The
behavior of the electric fields for ENd black holes and different
value of $\alpha$ is shown in figure \ref{figure2}. From this
figure we see that for $\alpha=0$, the electric field has a finite
value \emph{near} the origin, while it diverges exactly at $r=0$.
This is in contrast to the BI electrodynamics. Again, with
increasing $\alpha$, the divergency of the electric field
increases  near the origin where $r\rightarrow 0$. In figures 3
and 4 we have compared the behavior of $E(r)$ for BId, ENd and EMd
black holes for both $\alpha=0$ and $\alpha=0.4$. Finally, we have
plotted in figure 5 the electric field of ENd black holes for
different values of the nonlinear parameter $\beta$. From this
figure we see that with increasing $\beta$, the electric field
diverges near the origin. This is an expected result, since for
large $\beta$ our theory reduces to the well-known EMd gravity
\cite{Shey3}.
\begin{figure}[tbp]
\epsfxsize=7cm \centerline{\epsffile{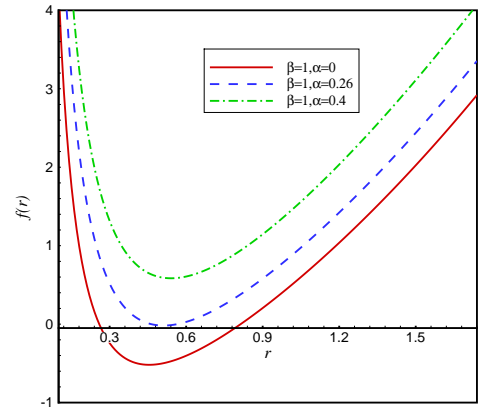}} \caption{ $f(r)$
versus $r$ for $q=1$ and  $m=2.5$.} \label{figure6}
\end{figure}
\begin{figure}[tbp]
\epsfxsize=7cm \centerline{\epsffile{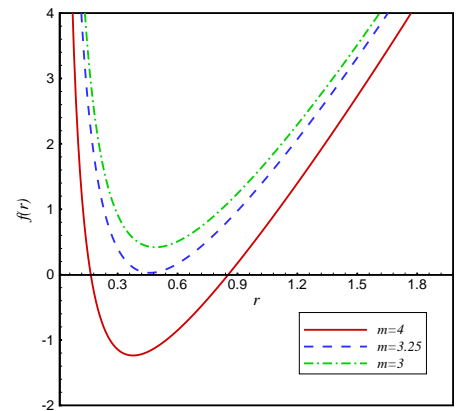}} \caption{$f(r)$
versus $r$ for $q=1$, $\protect\alpha=0.5$ and $\beta=1$.}
\label{figure7}
\end{figure}
\begin{figure}[tbp]
\epsfxsize=7cm \centerline{\epsffile{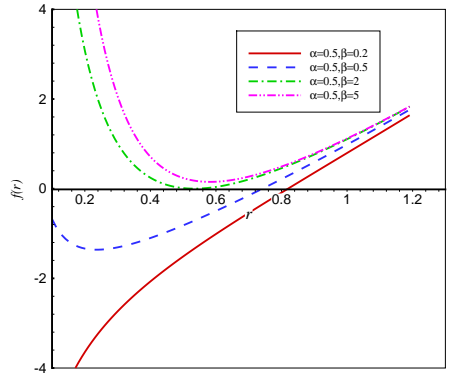}} \caption{$f(r)$
versus $r$ for $m=3.5$ and $\alpha=0.5$ and $q=1$.}
\label{figure8}
\end{figure}
\begin{figure}[tbp]
\epsfxsize=7cm \centerline{\epsffile{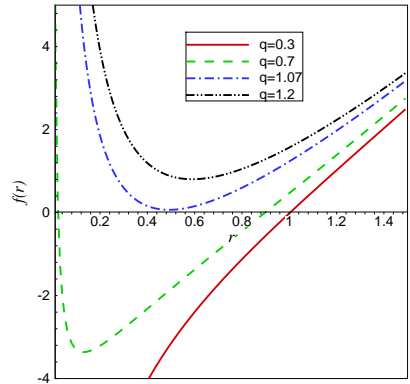}} \caption{$f(r)$
versus $r$ for $m=3.5$ and $\alpha=0.5$ and $\beta=1$.}
\label{figure9}
\end{figure}
\begin{figure}[tbp]
\epsfxsize=7cm \centerline{\epsffile{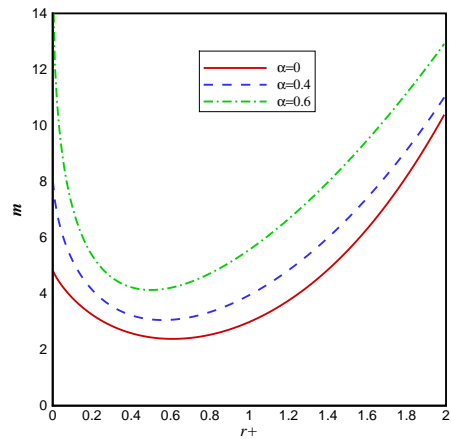}} \caption{The mass
parameter $m$ versus $r_{+}$ for $\beta=2$ and $q=1$. }
\label{figure10}
\end{figure}
\begin{figure}[tbp]
\epsfxsize=7cm \centerline{\epsffile{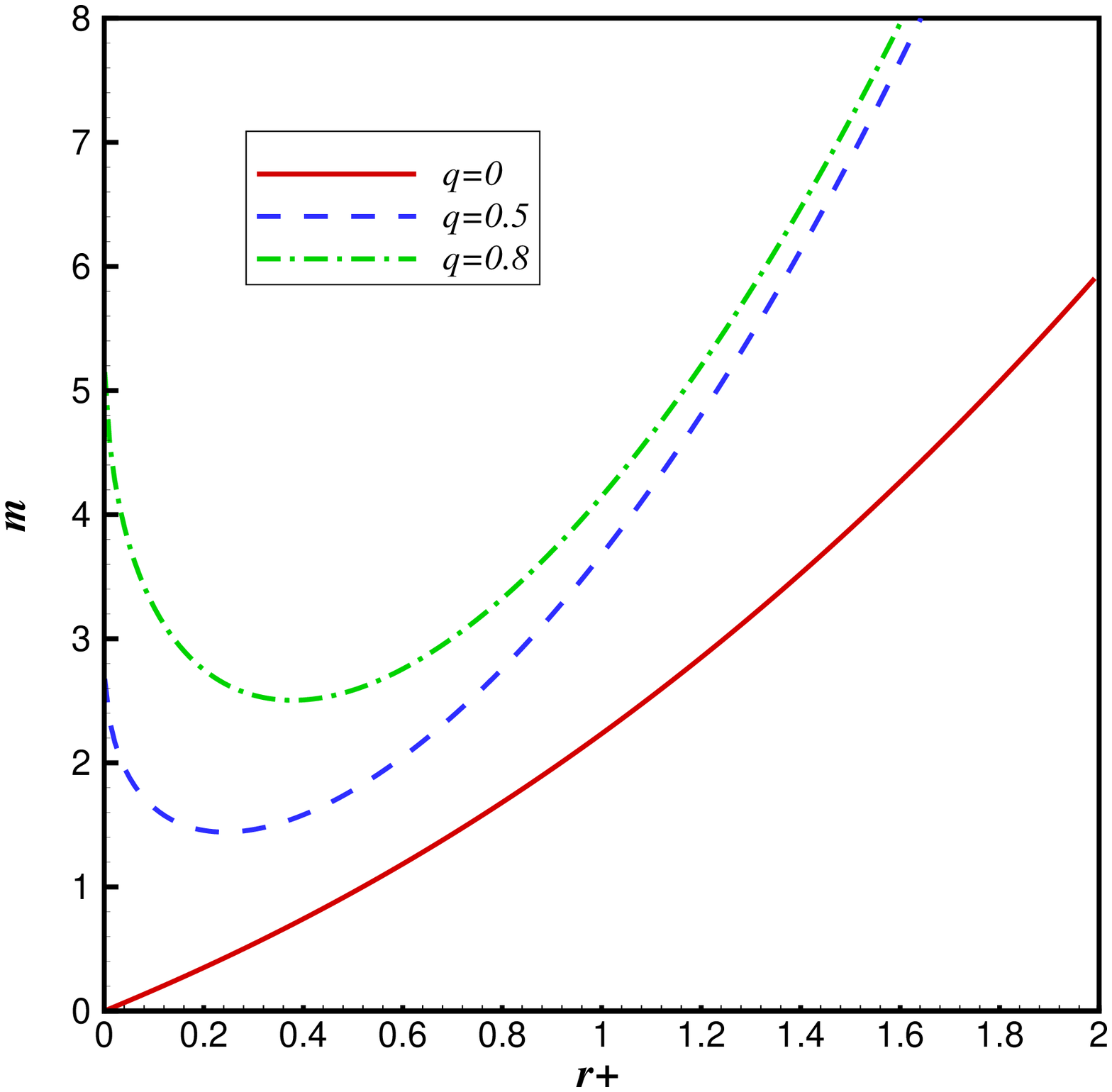}} \caption{The mass
parameter $m$ versus $r_{h}$ for  $\alpha=0.5$ and $\beta=1$.}
\label{figure11}
\end{figure}

Next, we look for the curvature singularities and horizons. In the
presence of the dilaton field, the Kretschmann scalar $R_{\mu \nu
\lambda \kappa }R^{\mu \nu \lambda \kappa }$ diverges as
$r\rightarrow 0$. Thus, the spacetime has an essential singularity
at $r=0$. However, the spacetime is neither asymptotically flat
nor (A)dS. As one can see from Eq. (\ref{f1}), the solution is
ill-defined for $\alpha^2 =3$ and $\alpha^2 =1$. Now, we search
for the horizons of spacetime. The horizons, if any exist can be
obtained by solving $f(r_{+})=0$. Clearly, due to the complexity
of $f(r)$ given in (\ref{f2}), it is not possible to find the
roots of $f(r_{+})=0$, analytically. However, we have plotted the
function $f(r)$ versus $r$ for different model parameters in
figures \ref{figure6}-\ref{figure9}. For simplicity, in these
figures, we kept fixed the other parameters $l=b=1$. These figures
show that the obtained solutions may represent a black hole with
two horizons, an extreme black hole or a naked singularity
depending on the metric parameters. For example, figure
\ref{figure6} shows that for fixed value of the other parameters,
the number of horizons decreases with increasing $\alpha$, while
one can see from figure \ref{figure7} that with increasing $m$,
the number of horizons increases. Figure \ref{figure8} shows that
for fixed value of $m$, $\alpha$ and $q$, there is a minimum
(extreme) value for the nonlinear parameter $\beta_{\rm min}$
$(\beta_{\rm ext})$, for which we have black hole with a
non-extreme horizon provided $\beta\leq\beta_{\rm min}$, black
hole with two horizons for $\beta_{\rm min}<\beta<\beta_{\rm
ext}$, black hole with an extreme horizon for $\beta=\beta_{\rm
ext}$ and naked singularity for $\beta>\beta_{\rm ext}$. Clearly,
$\beta_{\rm min}$ and $\beta_{\rm ext}$ depend on the other
parameters of the model. Similar, argument can be applied for the
charge parameter, $q$, which is shown in figure \ref{figure9}.

To have further understanding on the nature of the horizons, we
plot in figures \ref{figure10} and \ref{figure11}, the mass
parameter $m$ as a function of the horizon radius $r_{+}$ for
different model parameters. Again, we set $l=b=1$, for simplicity.
These figures show that, for fixed value of other parameters, the
value of $m$ determines the number of horizons. We see that, up to
a certain value of the mass parameter $m$, there are two horizons,
and as we decrease $m$ further, the two horizons meet. In this
case we get extremal black hole (see the next section). Using the
fact that $f(r_{+})=0$, we can obtain the mass parameter in terms
of the horizon radius,
 \begin{eqnarray}\label{mrh}
m(r_{+})&=&-\frac{ {\alpha}^{2}+1}{ {\alpha}^{2}-1
}{b}^{-\gamma}{r_{+}}+\frac{(\Lambda+2\beta^2) \left(
{\alpha}^{2}+1 \right) ^{2}{b}^{\gamma}}{\alpha^{2}-3
}r_{+}^{3-2\gamma}+2 \beta q{(\alpha^2+1)^2}{(\frac{\beta^2
b^{2\gamma}}{q^2})}^{\frac{1-\gamma}{2\gamma-4}}{L_W^{\frac{3-2\gamma}{2\gamma-4}}(\eta_{+})}
\nonumber
\\ &&\times\Bigg{\{}\frac{L_W^2(\eta_{+})}{\alpha^2+5}\digamma\left(\frac{\alpha^2+5}{4},\frac{\alpha^2+9}{4},\frac{\alpha^2-1}{4}L_W(\eta_{+})\right)
-
\frac{1}{\alpha^2-3}\digamma\left(\frac{\alpha^2-3}{4},\frac{\alpha^2+1}{4},\frac{\alpha^2-1}{4}L_W(\eta_{+})\right)
\Bigg{\}}, \nonumber\\
\end{eqnarray}
where $\eta_{+}=\eta(r=r_{+})$. Figures \ref{figure11} also shows
that in the limit $r_{+}\rightarrow0$ we have a nonzero value for
the mass parameter $m$. This is in contrast to the Schwarzschild
black holes in which mass parameter goes to zero as
$r_{+}\rightarrow0$. This is due to the effect of the nonlinearity
of the electrodynamic field and in case of $q=0$, the mass
parameter $m$ goes to zero as $r_{+}\rightarrow0$.
\section{Conserved and Thermodynamics Quantities} \label{Therm}
\begin{figure}[tbp]
\epsfxsize=7cm \centerline{\epsffile{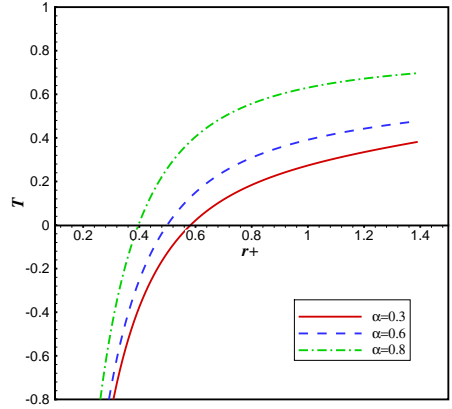}} \caption{ $T$
versus $r_{+}$ for $\beta=2$, $\Lambda=-3$ and $q=1$. }
\label{figure12}
\end{figure}

\begin{figure}[tbp]
\epsfxsize=7cm \centerline{\epsffile{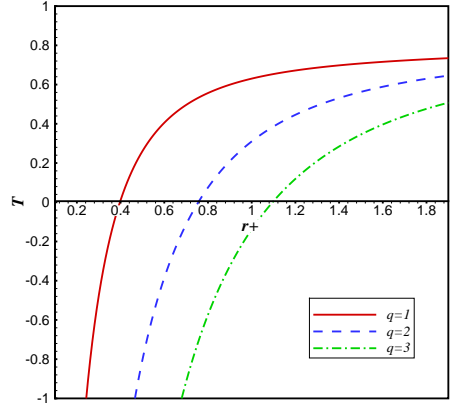}} \caption{$T$
versus $r_{+}$ for  $\alpha=0.8$,  $\Lambda=-3$ and $\beta=2$. }
\label{figure13}
\end{figure}

\begin{figure}[tbp]
\epsfxsize=7cm \centerline{\epsffile{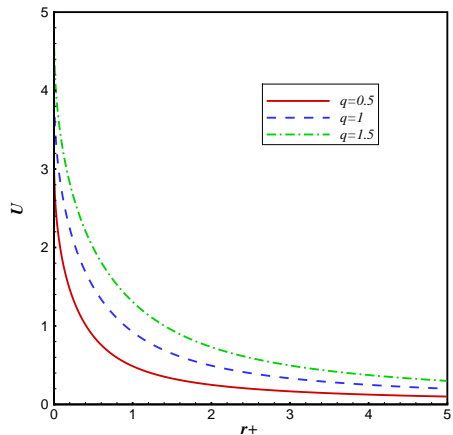}}
\caption{$U(r_{+})$ versus $r_{+}$ for  $\alpha=0.5$ and
$\beta=1$. } \label{figure14}
\end{figure}

\begin{figure}[tbp]
\epsfxsize=7cm \centerline{\epsffile{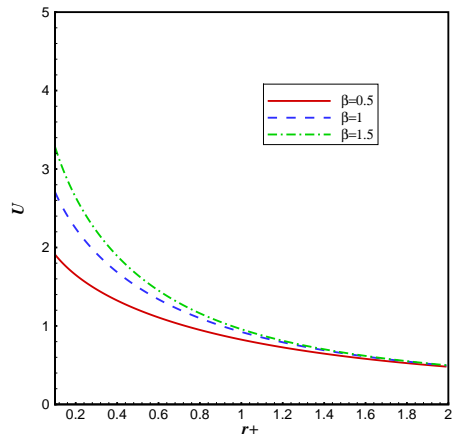}} \caption{
$U(r_{+})$ versus $r_{+}$ for $\alpha=0.5$ and $q=1$.}
\label{figure15}
\end{figure}

\begin{figure}[tbp]
\epsfxsize=7cm \centerline{\epsffile{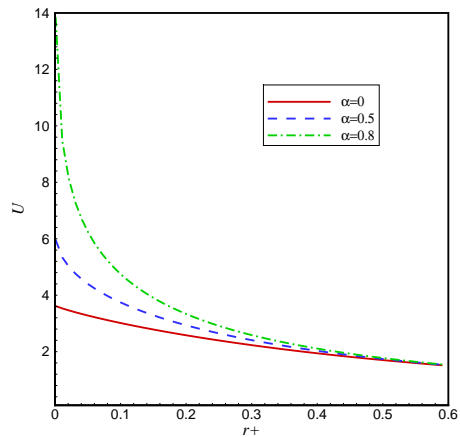}}
\caption{$U(r_{+})$ versus $r_{+}$ for  $q=1$ and $\beta=2$.}
\label{figure16}
\end{figure}

In this section we want to compute the conserved and thermodynamic
quantities of the ENd black hole solutions we just found. {There
are several ways for calculating the mass of the black holes. For
example, for asymptotically AdS solution one can use the
conterterm method inspired by (A)dS/CFT correspondence
\cite{Mal,Haro}. Another way for calculating the mass is through
the use of the substraction method of Brown and York \cite{BY}.
Such a procedure causes the resulting physical quantities to
depend on the choice of reference background. In our case, due to
the presence of the non-trivial dilaton field, the asymptotic
behaviour of the solutions are neither flat nor (A)dS, therefore
we have used the reference background metric and calculate the
mass. According to the substraction method of \cite{BY}, if we
write the metric of static spherically symmetric spacetime in the
form \cite{CHM}}
\begin{equation}\label{metric2}
ds^2=-W^2(r)dt^2 + {dr^2\over V^2(r)}+ r^2d\Omega^2,
\end{equation}
and the matter action contains no derivatives of the metric, then
the quasilocal mass is given by \cite{CHM}
\begin{equation}\label{QLM}
{\cal M} = r{W(r)}\left( {V_{0}(r)}- {V(r)}\right).
\end{equation}
Here $V_{0}(r)$ is an arbitrary function which determines the zero
of the energy for a background spacetime and $r$ is the radius of
the spacelike hypersurface boundary. It was argued that the $ADM$
mass $M$ is the ${\cal M}$ determined in (\ref{QLM}) in the limit
$r\rightarrow\infty$ \cite{CHM}. Transforming  metric
(\ref{metric}) in the form (\ref{metric2}), the mass of the ENd
black hole is obtained as
\begin{equation}
{M}=\frac{b^{\gamma}m \omega}{8\pi(\alpha^2+1)}.\label{Mass}
\end{equation}
where $\omega$ is the area of an unit $2$-sphere. {One can obtain
the temperature of the horizon by analytic continuation of the
metric. The analytical continuation of the Lorentzian metric by
$t\rightarrow i\tau $
yields the Euclidean section, whose regularity at $%
r=r_{+}$ requires that we should identify $\tau \sim \tau +\beta
_{+}$, where the period $\beta _{+}=1/T$ is the inverse Hawking
temperature $T$ of the horizon. It is a matter of calculation to
show that}
\begin{eqnarray}\label{Tem}
T_{+}&=&\frac{1}{4\pi}\left(\frac{df(r)}{dr}\right)_{r=r_{+}}=-\frac{(\alpha^2+1)}{4\pi}{r_{+}
^{1-\gamma}}\Bigg{\{}\frac{b^{-\gamma}
r_{+}^{2\gamma-2}}{\alpha^2-1}+(\Lambda+2\beta^2)b^{\gamma}-2\beta
q
r_{+}^{\gamma-2}\left(\frac{1}{\sqrt{L_{W}(\eta_{+})}}-\sqrt{L_{W}(\eta_{+})}\right)\Bigg{\}},
\nonumber\\
\end{eqnarray}
where $\eta_{+}=\eta(r=r_{+})$ and we have used $f(r_{+})=0$. The
behavior of $T$ versus $r_{+}$ is shown in figures \ref{figure12}
and \ref{figure13}. From these figures we find out that, for large
value of $r_{+}$, the temperature tends to a constant independent
of the model parameters. On the other hand, for small values of
$r_{+}$, the temperature may be negative ($T < 0$). In this case
we encounter a naked singularity. The temperature is zero and the
horizon is degenerate for an extremal black hole. In this case
$r_{\rm ext}$ is the positive root of the following equation:
\begin{eqnarray}\label{rext}
\frac{b^{-2\gamma}
r_{\rm{ext}}^{2\gamma-2}}{\alpha^2-1}+\Lambda+2\beta^2-2\beta
q_{\rm ext}
b^{-\gamma}r_{\mathrm{ext}}^{\gamma-2}\left(\frac{1}{\sqrt{L_{W}(\eta_{\rm
ext})}}-\sqrt{L_{W}(\eta_{\rm ext})}\right)=0.
\end{eqnarray}
where
\begin{equation}
\eta_{\rm ext}\equiv \frac{q_{\rm ext}^{2}r_{\rm
ext}^{2\gamma-4}}{\beta ^{2}b^{2\gamma }}. \label{etaext}
\end{equation}
From figures \ref{figure12}  and \ref{figure13}, we see that
$r_{\rm ext}$ decreases as $\alpha$ increases, while $r_{\rm ext}$
increases with increasing $q$. Indeed, the metric of Eqs.
(\ref{metric}) and (\ref{f2}) can describe a nonlinear dilaton
black hole with inner and outer event horizons located at $r_{-}$
and $r_{+}$, provided $r>r_{\mathrm{ext}}$, an extreme ENd black
hole in case of $r=r_{\mathrm{ext}}$, and a naked singularity if
$r<r_{\mathrm{ext}}$. Note that in the limiting case where
$\beta\rightarrow\infty$, expression (\ref{Tem}) reduces to that
of black hole in EMd theory \cite{Shey3}
\begin{eqnarray}\label{Tem2}
T_{+}&=&-\frac{b^{-\gamma}(\alpha ^2+1)}{8
\pi(\alpha^2-1)}r_{+}^{\gamma-1}-\frac{\Lambda (\alpha ^2+1)
b^{\gamma}}{4 \pi}r_{+}^{1-\gamma}-\frac{q^{2}b^{-\gamma} (\alpha
^2+1)}{4 \pi}r_{+}^{\gamma-3}\nonumber \\
&&+\frac{q^{4}b^{-3\gamma} (\alpha ^2+1)}{16 \pi
\beta^2}r_{+}^{3\gamma-7}+O \left(\frac{1}{\beta^4}\right).
\end{eqnarray}
{The entropy of the black hole satisfies the so called area law of
the entropy which states that the entropy of the black hole is a
quarter of the event horizon area \cite{Beck}. This near universal
law applies to almost all kinds of black holes, including dilaton
black holes, in Einstein gravity \cite{hunt}.}
 It is easy to show
\begin{equation}
{S}=\frac{A}{4}=\frac{
b^{\gamma}r_{+}^{2-\gamma}\omega}{4}.\label{Entropy}
\end{equation}
Using the Gauss's law, we can calculate the flux of the
electromagnetic field at infinity to obtain the electric charge of
black hole as
\begin{equation}
{Q}= \frac{1}{4\pi} \int e^{-2\alpha\Phi}\text{ }^{*} F
d{\Omega}=\frac{q \omega}{4\pi}.  \label{Q}
\end{equation}%
The electric potential $U$, measured at infinity with respect to
the horizon, is defined by
\begin{equation}
U=A_{\mu }\chi ^{\mu }\left| _{r\rightarrow \infty }-A_{\mu }\chi
^{\mu }\right| _{r=r_{+}},  \label{Pot}
\end{equation}
where $\chi=\partial_{t}$ is the null generator of the horizon.
The gauge potential $A_{t }$ corresponding to the electromagnetic
field (\ref{FtrE2}) can be written as
\begin{eqnarray}\label{vectorpot}
A_{t}&=&b^{\gamma}\beta(\alpha^2+1)\left(\frac{\beta
b^{\gamma}}{q}\right)^{\frac{1-\gamma}{\gamma-2}}\left(\frac{1-\alpha^2}{4}\right)^{\frac{1}{2\gamma-4}}\nonumber
\\ && \times \Bigg{\{}
-\frac{1}{4}\Gamma\left(\frac{\alpha^2+1}{4},\frac{1-\alpha^2}{4}L_{W}(\eta)\right)
+\frac{1}{\alpha^2-1}\left[\Gamma\left(\frac{\alpha^2+5}{4},\frac{1-\alpha^2}{4}L_{W}(\eta)\right)-\frac{1}{2}\Gamma\left(\frac{\alpha^2+1}{4}\right)\right]
\Bigg{\}}.\nonumber \\
\end{eqnarray}
Therefore, the electric potential may be obtained as
\begin{eqnarray}\label{Pot}
U&=&b^{\gamma}\beta(\alpha^2+1)\left(\frac{\beta
b^{\gamma}}{q}\right)^{\frac{1-\gamma}{\gamma-2}}\left(\frac{1-\alpha^2}{4}\right)^{\frac{1}{2\gamma-4}}\nonumber
\\ && \times\Bigg{\{}
-\frac{1}{4}\Gamma\left(\frac{\alpha^2+1}{4},\frac{1-\alpha^2}{4}L_{W}(\eta_{+})\right)
+\frac{1}{\alpha^2-1}\left[\Gamma\left(\frac{\alpha^2+5}{4},\frac{1-\alpha^2}{4}L_{W}(\eta_{+})\right)-\frac{1}{2}\Gamma\left(\frac{\alpha^2+1}{4}\right)\right]
\Bigg{\}}.\nonumber \\
\end{eqnarray}
Expanding for large value of $\beta$, we get
\begin{eqnarray}
U=\frac{q}{r_{+}}-\frac{ b^{-2\gamma}(\alpha^2+1) }{2\beta^2
(\alpha^2+5)
}\frac{q^3}{r_{+}^{5-2\gamma}}+O\left(\frac{1}{\beta^4}\right).
\end{eqnarray}
We have shown the the behavior of the electric potential $U$ as a
function of horizon radius $r_{+}$ in figures
\ref{figure14}-\ref{figure16} for $b=1$. Due to the nature of the
nonlinear electrodynamics, the electric potential can be finite as
$r_{+}\rightarrow0$, depending on the model parameters, and goes
to zero for large $r_{+}$ independent of the model parameters.
From these figures we find that for fixed value of other
parameter, the divergency of $U$, for small $r_{+}$, increases
with increasing $\alpha$ and $\beta$.

Having the conserved and thermodynamic quantities at hand, we are
in a position to check the first law of thermodynamics for the
obtained solutions. For this purpose, we first obtain the mass $M$
as a function of extensive quantities $S$ and $Q$. Combining
expressions for the charge, the mass and the entropy given in Eqs.
(\ref{Mass}), (\ref{Entropy}) and (\ref{Q}), and using the fact
that $f(r_{+})=0$, we obtain a Smarr-type formula as
\begin{eqnarray}\label{Msmar1}
M(S,Q)&=&-\frac{ b^{-\alpha^2}{\left(4S\right)}^{(\alpha^2+1)/2}
}{8\pi(\alpha^2-1)} +\frac{(\alpha^2+1)b^{\alpha^2}}{8\pi
(\alpha^2-3)}{(\Lambda+2\beta^2)}
{\left(4S\right)}^{(3-\alpha^2)/2}
\nonumber\\
&&+\frac{b^\gamma\beta
Q}{4}\frac{(\alpha^2+1)^3}{(\alpha^2-1)^2}{\left(\frac{\beta^2
b^{2\gamma}}{16\pi^2
Q^2}\right)}^{\frac{1-\gamma}{2\gamma-4}}\left(\frac{1-\alpha^2}{4}\right)^{\frac{2\gamma-3}{2\gamma-4}}\nonumber
\\ && \times \Bigg{\{} -4(\gamma-2)^2\left[\Gamma
\left(\frac{\alpha^2+5}{4},\frac{1-\alpha^2}{4}L_{W}\left(\zeta\right)\right)-\Gamma\left(\frac{\alpha^2+5}{4}\right)\right]\nonumber \\
&&+(\gamma-1)^2\left[\Gamma
\left(\frac{\alpha^2-3}{4},\frac{1-\alpha^2}{4}L_{W}\left(\zeta\right)\right)
-\Gamma\left({\frac{\alpha^2-3}{4}}\right)\right]\Bigg{\}},
 \end{eqnarray}
where $\zeta=\frac{\pi^2Q^2}{S^2\beta^2}$. If we expand $M(S,Q)$
for large $\beta$, we arrive at
\begin{eqnarray}
M(S,Q)&=&-\frac{ b^{-\alpha^2}{\left(4S\right)}^{(\alpha^2+1)/2}
}{8\pi(\alpha^2-1)} +\frac{\Lambda(\alpha^2+1)b^{\alpha^2}}{8\pi
(\alpha^2-3)}{\left(4S\right)}^{(3-\alpha^2)/2}\nonumber \\&& +2
\pi Q^2 b^{\alpha^2}(4 S)^{(-\alpha^2-1)/2}-\frac{8 Q^4 \pi^3
(\alpha^2+1)b^{\alpha^2}}{\beta^2(\alpha^2+5)}
(4S)^{(-\alpha^2-5)/2}+O\left(\frac{1}{\beta^4}\right),
\label{Msmar2}
 \end{eqnarray}
which is exactly the Smarr-type formula obtained for EMd black
holes in the limit $\beta^2\rightarrow\infty$ \cite{Shey3}. Now,
if we consider $S$ and $Q$ as a complete set of extensive
parameters for the mass $M(S,Q)$, we can define the intensive
parameters conjugate to $S$ and $Q$ as
\begin{equation}
T=\left( \frac{\partial M}{\partial S}\right) _{Q},\ \ U=\left( \frac{\partial M%
}{\partial Q}\right) _{S}.  \label{Dsmar}
\end{equation}
After numerical calculations we can show that the intensive
quantities calculated by Eq. (\ref{Dsmar}) coincide with Eqs.
(\ref{Tem}) and (\ref{Pot}). Thus, these thermodynamic quantities
satisfy the first law of black hole thermodynamics
\begin{equation}
dM = TdS+Ud{Q}.
\end{equation}
{The satisfaction of the first law of thermodynamics for the
obtained conserved and thermodynamic quantities, together with the
fact that these quantities in two limiting cases, namely in the
absence of the dilaton field ($\alpha=0=\gamma$), and for large
values of the nonlinear parameter ($\beta\rightarrow\infty$),
reduce to the known results in the literature \cite{Hendi2,Shey3},
indicate that the conserved and thermodynamic quantities obtained
in this paper are correct and in agreement with other method such
as Euclidean action method \cite{Mis}.}
\section{closing remarks}
The pioneering theory of the nonlinear electromagnetic field was
proposed by Born and Infeld in $1934$ for the purpose of solving
various problems of divergence appearing in the Maxwell theory
\cite{BI}. However, in recent years, other modifications of linear
Maxwell's field have been proposed which yield finite self-energy
for a charged point particle located at $r=0$. Among them, are
exponential and logarithmic form of Lagrangian for nonlinear
electrodynamics. Since the series expansion and the behavior of
the solutions of these two Lagrangian, for large value of the
nonlinear parameter, are the same as BI Lagrangian, they usually
called BI-like Lagrangian in the literatures \cite{Hendi2}. Black
hole solutions in the presence of BI-like nonlinear
electrodynamics have been investigated in
\cite{Soleng,HendiJHEP,Hendi2}.

In this paper, we have extended the study on the EN
electrodynamics by taking into account the dilaton scalar field in
the action. We first proposed the suitable Lagrangian for EN
electrodynamics coupled to the dilaton field. As far as we know,
this is for the first time which Lagrangian (\ref{LFPhi}) is
introduced. We have compared in Eqs. (8) and (9) the series
expansions of ENd Lagrangian with BId Lagrangian in the limit of
large $\beta$, and found that they are the same. We have varied
the action and obtained the field equations of ENd theory. Then,
by making a suitable ansatz (\ref{Rphi}), we have constructed a
new class of charged static and spherically symmetric black hole
solutions in the presence of ENd electrodynamics. In the limiting
case where $\beta\rightarrow\infty$, our solutions reduce to EMd
black hole solutions \cite{Shey3}, while in the absence of the
dilaton field, ($\alpha=0=\gamma$), they restore charged black
holes coupled to EN electrodynamics \cite{Hendi2}. We also
investigated and compared the behavior of the electric field of
three kind of black holes in dilaton gravity, namely EMd, BId and
ENd black holes we just found. Although the behavior of the
electric fields near the origin depends on the model parameters,
however for large $r$ the asymptotic behavior of all of them are
exactly the same as linear Maxwell field. Interestingly enough, we
found that the electric fields of ENd black hole is finite near
the origin and diverges exactly at $r=0$ depending on the model
parameters, however its divergency is much slower than the Maxwell
field.

We also investigated the physical properties of the solutions in
ample details. The presence of the dilaton field changes the
asymptotic behavior of the solutions to be neither flat nor (A)dS.
Our solutions can represent black holes with inner and outer
horizons, an extreme black hole or naked singularity depending on
the model parameters. For fixed value of $m$, $\alpha$ and $q$, we
found that there is a minimum (extreme) value for nonlinear
parameter $\beta_{\rm min}$ $(\beta_{\rm ext})$, for which we have
black hole with a non-extreme horizon provided
$\beta\leq\beta_{\rm min}$, black hole with two horizons for
$\beta_{\rm min}<\beta<\beta_{\rm ext}$, black hole with an
extreme horizon for $\beta=\beta_{\rm ext}$ and naked singularity
for $\beta>\beta_{\rm ext}$. We computed the mass, entropy,
temperature, electric potential of these black holes. We obtained
the Smarr-type formula,$M(S,Q)$, and checked that the conserved
and thermodynamic quantities obtained for these solutions satisfy
the first law of black holes thermodynamics on the horizon.

Finally, we would like to mention that black hole solutions we
obtained here are static. Thus, it would be nice to derive
rotating black hole/string solutions of these field equations.
Besides, it is also interesting to generalize the study to higher
dimensions and construct both static and rotating black
holes/branes of ENd theory in arbitrary dimensions. In addition,
we only considered the exponential nonlinear electrodynamics
coupled to the dilaton field, so the case with logarithmic
nonlinear electrodynamics in the presence of dilaton field remains
to be investigated. Our group are now working on these subjects
and the results will be appeared shortly in our future works.

\acknowledgments{We thank referee for constructive comments which
helped us to improve the paper significantly. A. Sheykhi thanks
Shiraz University Research Council. This work has been supported
financially by Research Institute for Astronomy and Astrophysics
of Maragha, Iran.}

\end{document}